# Plasma Profiling Time-of-Flight Mass Spectrometry for Fast Elemental Analysis of Semiconductor Structures with Depth Resolution in the Nanometer Range

Hendrik Spende[1,2], Christoph Margenfeld[1,2], Tobias Meyer[3], Irene Manglano Clavero[1,2], Heiko Bremers[2,4], Andreas Hangleiter[2,4], Michael Seibt[3], Andreas Waag[1,2], Andrey Bakin[1,2]

[1] Institute of Semiconductor Technology, Technische Universität Braunschweig, Hans-Sommer-Straße 66, D-38106 Braunschweig, Germany
[2] Laboratory for Emerging Nanometrology (LENA), Technische Universität Braunschweig, Langer Kamp 6, D-38106 Braunschweig, Germany
[3] 4th Physical Institute – Solids and Nanostructures, Georg-August-University Göttingen, Friedrich-Hund-Platz 1, D-37077 Göttingen, Germany
[4] Institute of Applied Physics, Technische Universität Braunschweig, Mendelssohnstr. 2, D-38106 Braunschweig, Germany

E-mail: h.spende@tu-braunschweig.de

**Abstract**

Plasma profiling time of flight mass spectrometry (PP-TOFMS) has recently gained interest as it enables the elemental profiling of semiconductor structures with high depth resolution in short acquisition times. As recently shown by Tempez et al., PP-TOFMS can be used to obtain the composition in the structures for modern field effect transistors [1]. There, the results were compared to conventional SIMS measurements. In the present study, we compare PP-TOFMS measurements of an Al-/In-/GaN quantum well multi stack to established micro- and nanoanalysis techniques like cathodoluminescence (CL), scanning transmission electron microscopy (STEM), energy dispersive X-ray spectroscopy (EDX) and X-ray diffraction (XRD). We show that PP-TOFMS is able to resolve the layer structure of the sample even more than 500 nm deep into the sample and allows the determination of a relative elemental composition with an accuracy of about 10 rel. %. Therefore, it is an extremely rapid alternative method to obtain semiconductor elemental depth profiles without the expensive and time consuming sample preparation required for TEM. Besides, PP-TOFMS offers better depth resolution and more elemental information than, for example, electrochemical capacitance-voltage (ECV), since all elements are detected in parallel and not only electrically (ECV) or optically (CL) active elements are observed.

Keywords: PP-TOFMS, XRD, TEM, GaN, InGaN, AlGaN, MQW

## 1. Introduction

The complexity of semiconductor structures is increasing with technological progress and established characterization methods are beginning to struggle with the intricacy of the structures. For example, X-ray diffraction (XRD) data in multi-layer structures are difficult to interpret, as these structures can cause many peaks, some of which partly overlap. Time-of-flight secondary ion mass spectroscopy (TOF-SIMS) and various techniques of transmission electron microscopy (TEM), such as energy dispersive X-ray spectroscopy (EDX) or electron energy loss spectroscopy (EELS), are alternative tools for the analysis of sample structure and elemental composition. But all these techniques are very costly and the interpretation of the data is complex and time consuming, thus requiring extremely highly qualified personnel. PP-TOFMS is a complementary and rather new method to obtain elementary depth profiles in a short time. In recent studies PP-TOFMS profiles have been compared to secondary ion mass spectroscopy (SIMS) and PP-TOFMS has provided semi-quantitative data of semiconductor hetero-structures and contact structures for transistors [1,2].

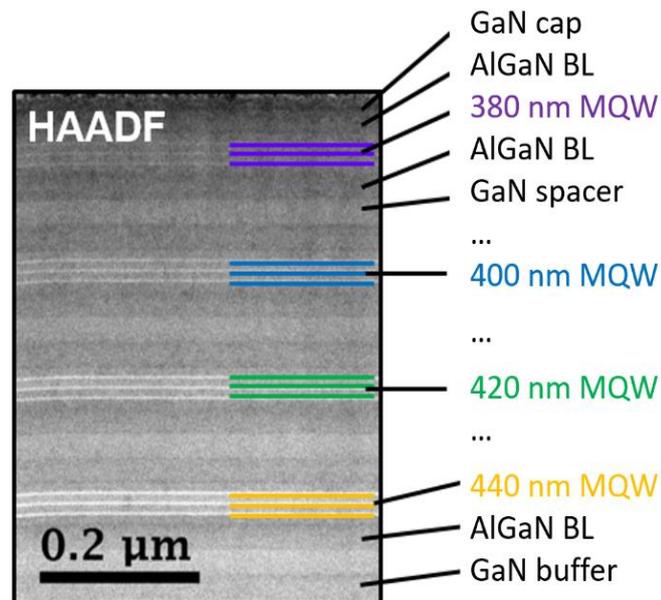

**Figure 1.** *STEM-HAADF micrograph (300 kV) showing the Al-/In-/GaN sample structure. The different layers are marked and the overlay highlights the four threefold MQWs, each emitting at a different wavelength.*

In this study, we compare PP-TOFMS data of an even more complex Al-/In-/GaN hetero-structure to well established elemental analysis tools such as XRD and STEM-EDX. CL emission and structure simulations provide access to the indium content in the optically active structures. The results are discussed in terms of depth resolution and composition.

## 2. Sample description

This study examines an InAlGaN sample with four InGaN/GaN multi-quantum-wells (MQWs), each designed for a different emission wavelength and stacked in growth direction, to obtain information in the nanometer range even at depths of 500 nm and more. A micrograph of the structure is shown in Figure 1.

The sample was grown employing metal organic vapor phase epitaxy (MOVPE) inside an industrial scale Aixtron 2600 HT G3 planetary reactor on 430 μm thick 2" c-plane sapphire substrates using standard precursors trimethylgallium (TMGa), triethylgallium (TEGa), trimethylindium (TMIn), trimethylaluminium (TMAl), monosilane ($SiH_4$) and ammonia ($NH_3$), as well as hydrogen and nitrogen as carrier gases. The growth process consists of a thermal cleaning of the sapphire wafers at 1100 °C in hydrogen atmosphere, the deposition of a low-temperature GaN nucleation layer, a thermal recrystallization step and coalescence of the nuclei by further GaN growth at 1050 °C. In total, a moderately silicon-doped (electron concentration of approx. $1 \times 10^{18}$ cm$^{-3}$) GaN buffer layer of 4 μm thickness was deposited at a reactor pressure of 290 hPa and a V/III ($NH_3$/TMGa) ratio of around 1000 using hydrogen as carrier gas. Following the buffer layer growth, four different active regions were grown, each consisting of a threefold MQW in order to increase the scattering cross section for electrons during CL experiments. Each MQW stack is sandwiched between approx. 30 nm thick AlGaN carrier blocking layers to avoid diffusion of charge carriers between the MQWs and separated by a 26 nm thick GaN spacer grown at 960 °C, 50 hPa reactor pressure and a TMAl/(TMAl+TMGa) inlet ratio of 0.15 for the AlGaN layer. The active regions are grown at low temperature using TEGa and TMIn as group III precursors, using a high V/III ratio of 22600 and a reactor pressure of 200 hPa. The MQW stack consists of 9 nm thick quantum barriers grown at 795 °C and 3 nm thick quantum wells deposited with a TMIn/(TMIn+TEGa) inlet ratio of 0.45, both using nitrogen as carrier gas. The indium incorporation into the InGaN films is thermodynamically limited under the applied growth conditions and thus can be tuned via the growth temperature [3,4]. To reduce the reabsorption of the optical emission in overlaying quantum wells, the indium content was decreased (and respectively the band gap of the MQWs increased) in growth direction by choosing growth temperatures of 705, 720, 735 and 750 °C, respectively. In order to exclude the influence of electric fields (e.g. due to different doping) the structure was not intentionally doped following the growth of the buffer layer and hence is slightly n-conductive.

**3. Methods**

*3.1 PP-TOFMS*

The PP-TOFMS by Horiba Scientific is an elementary depth profiler based on glow discharge plasma sputtering and time-of-flight mass spectrometry. The sample is sputtered by argon ions created in a pulsed radio frequency plasma and the atomic species of the sample are ionized in the plasma and detected by an orthogonal time-of-flight mass spectrometer. The method is described in detail elsewhere [5,6].

In the PP-TOFMS by Horiba Scientific, the samples are placed vacuum-tight "face down" on a 4 mm diameter sealing-ring on the anode. Sample and source are simultaneously pumped and purged with 6N argon gas at a constant pressure of 300 Pa in the source for 180 seconds in order to reduce the remaining residual atmosphere species. The main plasma parameters – 6N argon pressure of 140 Pa to 150 Pa and RF powers of 50 and 55 W – were optimized for our samples in order to resolve thin layers as precise as possible. Plasma auto-matching was used to achieve stable sputtering conditions. The pulse period was set to 550 µs at a pulse width of 150 µs, 100 pulses were averaged. A typical surface profile of the resulting crater is presented in Figure 2(b). The duration of the plasma profiling can be tuned to the measurement objective since all necessary ion profiles are monitored in real time during the measurement. Typical sputtering rates range from slightly below 1 nm/s to a few tens of nm/s. Thus, the entire measurement procedure of the sample up to the depth of approx. 500 nm with high depth resolution can be realized in 1 to 3 minutes.

The elemental contents of indium, aluminum and gallium were obtained using the ion beam ratio (IBR) method [7]. The IBR is calculated directly from the number of counts for each element x $\#_x$ divided by the sum of the counts for all group III metals $\#_m$:

$$IBR_x = \frac{\#_x}{\sum_{m=1}^{\infty} \#_m}$$

A major advantage of the IBR method is that plasma instabilities do not affect the result, since the normalization only takes into account the ratio between different elements. Therefore, it is assumed that the composition of the plasma represents the composition of the sputtered region, which is applicable as the ionization yield for the elements is constant for a wide range of atomic masses [8]. This assumption does not hold for some elements such as nitrogen or oxygen as they are difficult to ionize and would hence be underrepresented in the IBR [8]. The counts in this calculation are not corrected for different sensitivities between the elements, which makes the IBR only an approximation of the elemental composition. However, since all In-/Al-/GaN species in the sample contain the same amount of nitrogen, nitrogen can be neglected in the IBR calculation and a relative composition may be obtained with sufficient accuracy.

Higher accuracy can be achieved by correcting for different sensitivities using relative sensitivity factors $RSF_x$ for each element x. The concentration of the element $c_x$ is then calculated by

$$c_x = \frac{RSF_x \, IBR_x}{\sum_{m=1}^{\infty} RSF_m \, IBR_m} \; .$$

RSFs are depending on element matrices and plasma conditions [9] and the determination of correct RSFs requires reliably calibrated reference materials (CRM). In pulsed mode, however, the RSF dependence on gas flow is reduced compared to continuous plasma modes and standard RSFs might be applicable. Unfortunately, reliable standard RSFs such as those available for metals [10], cannot be found for the nitride material system. In this study RSFs for Al, Ga and In are determined within a special reference measurement of the described sample and are compared to other elemental analysis tools like EDX and CL. The resulting RSFs were then used for the concentration determination of the elements during different measurement sessions.

*3.2 Cathodoluminescence*

Cathodoluminescence (CL) in a scanning electron microscope (SEM) was carried out with a Tescan Mira 3 GMH field emission SEM equipped with a Gatan Mono CL 4 setup with a parabolic mirror for light collection. The CL setup is equipped with a photo-multiplier-tube (PMT) for pan- and monochromatic CL-maps and an Andor DV420-BV charge-coupled device (CCD) for the acquisition of emission spectra. The 300 l/mm grating inside the monochromator offers a spectral resolution of ~3 nm on the CCD.

In principle, the indium content x and the band gap of a quantum well might be estimated from the emission energy as follows:

$$E_g = 3.4\,(1-x) + 0.77\,x - 1.43\,x\,(1-x)$$

With the band gap energies of GaN $E_{GaN} = 3.4$ eV and of InN $E_{InN} = 0.77$ eV as well as the bowing factor for InGaN $E_{InGaN} = 1.43$ eV [11].

Since this formula does not take into account strain and quantum effects, it only applies for thick InGaN layers and not to QWs as investigated in this study. The determination of elemental compositions from optical emission of quantum wells in nitrides is difficult because spontaneous and piezoelectric polarization charges at the quantum well interfaces cause a band inclination, the so-called quantum confined Stark effect (QCSE). Quantum effects also play a role. For a known layer stack, as in this study, it is possible to simulate the band structure. We choose Nextnano++ to simulate single InGaN QWs with a thickness of 2.75 nm to determine the indium content corresponding to the optical emission energy of each MQW. Nextnano++ [12] is a Schrödinger-Poisson drift-diffusion solver and a valuable tool for the simulation of band structures in semiconductor devices considering strain and quantization effects. The simulation does not consider the compositional inhomogeneity and thickness variations inside the InGaN QWs [13–15], which influence the optical emission of the QWs.

### 3.3 STEM

The TEM lamella was prepared in a FEI Nova NanoLab Dual Beam using a gallium ion acceleration voltage of 5 kV for the final thinning step. Subsequently, STEM measurements were carried out in an FEI Titan 80-300 operated at 300 kV for obtaining images and at 80 kV for reducing beam damage during EDX acquisitions. The acceptance angle of the high-angle annular dark-field detector was 55-200 mrad. Energy-dispersive X-ray data were collected with an Oxford

Instruments X-Max 80 mm² detector as follows. Using a beam current of 120 pA, the STEM probe was scanned across a 2D point array and integrated along the direction perpendicular to the growth direction, i.e. the second axis in this section, in order to reduce the dose on the sample. For
each MQW, a 50×10 px array was recorded with a spacing of 1.1 nm resp. 21.81 nm. In addition, an overview scan was acquired containing 350×4 px with a spacing of 1.29 nm resp. 27.72 nm. After integration along the second axis, HyperSpy v1.4.2 [16] was used to quantify the data following the Cliff-Lorimer method [17] with the set of k values ($k_{AlK\alpha} = 1.022$, $k_{GaK\alpha} = 2.249$, $k_{InL\alpha} = 2.742$). A background subtraction was performed for each line individually and both the integration and background windows of all lines are summarized in table 1.

| Line | Background window (pre line) [keV] | Background window (post line) [keV] | Integration window [keV] |
|---|---|---|---|
| AlKα | 1.3-1.4 | 2.3-2.5 | 1.4-1.6 |
| GaKα | 8.35-8.45 | 9.65-9.85 | 9.0-9.5 |
| InLα | 2.3-2.5 | 4.0-4.2 | 3.2-3.8 |

**Table 1.** *Energy windows used during the EDX analysis.*

### 3.4 High-Resolution X-Ray diffraction

High-resolution X-ray diffraction measurements are a standard technique for precise compositional analysis of epitaxially grown III-nitride thin films [18]. Rocking curves (ω-scans), coupled scans (ω-2θ-scans) and reciprocal space maps (RSMs) of several symmetric and asymmetric reflexes were recorded using a Panalytical X'Pert PRO system. The structural properties of the sample are determined by comparing simulations of a highly resolved triple-axis ω-2θ-scan of the (0002) reflex to measured data.

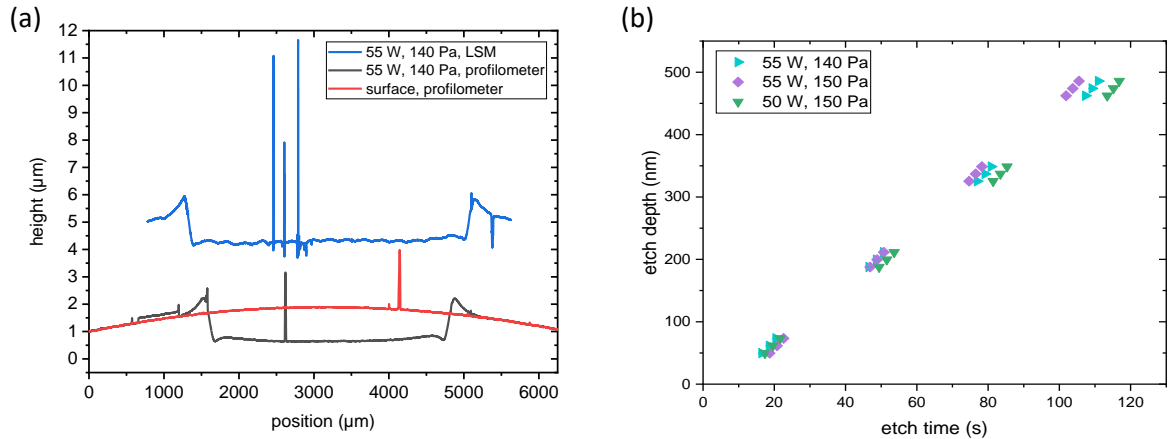

**Figure 2.** *(a) Laser scanning microscope and mechanical profilometer topographies of a crater after PP-TOFMS measurement and the wafer surface, (b) Correlation of the indium peaks in the TOF-MS signal (x-axis) and the bright contrast in the HAADF signal revealing the etch rate for different plasma parameters.*

**4. Results and Discussion**

*4.1 Etch/sputtering rate*

The easiest way to determine the sputtering rate is to measure the depth of the crater caused by the plasma and compare it with the measurement time in the PP-TOFMS. However, the large scale of the crater, several orders of magnitude larger than the etching depth, and the wafer bowing make precise measurement of the etching depth difficult. Uneven crater topographies and precipitation at the crater edge impede this even more. Figure 2(a) shows mechanical profilometer and laser scanning microscope (LSM) depth profiles of the crater and the wafer surface measured in close vicinity of the crater. The waviness on the crater bottom of the LSM measurement is due to the stitching of separate measurements. The peaks originate from spikes inside the crater which form during plasma ablation of the material.

Taking wafer bowing into account, the etch rates are in the range of 4.6 nm/s (50W, 150 Pa) to 5.5 nm/s (55W, 140 Pa) for the center of the crater and in the range of 2.8 nm/s (50W, 150 Pa) and 4.2 nm/s (55W, 140 Pa) for the rim of the crater.

As the structure of the sample was known precisely by TEM measurements, we were able to correlate the QW position of the TEM measurements with the indium peaks of the PP-TOFMS data to analyze the etch rate during the experiment. A linear regression shows a linear relation between etch time and etch depth as the MQWs line up on the curve. The etch rate obtained by the fit is 4.3–5.5 nm/s. This corresponds well with the estimation of the depth vs. time in the crater center. A closer look on the regression reveals, that the etch rate in the different MQW regions is higher than the linear regression. The etch rate rises with increasing indium content of the MQWs from 5.8–6.0 nm/s for the topmost MQW up to 6.5–6.9 nm/s for the MQW with the highest indium content. As the etch rates for the quantum well region are higher than the average, the etch rate of the AlGaN blocking layer must be smaller. This can be seen in the fluctuation of the total number of PP-TOFMS counts, which increases with indium content and decreases for higher aluminium content. This observation corresponds well to literature [19].

The u-shaped profile of the crater, which rises slightly to the rim, explains the steep flank of the in signal in the rise and the increasing tailing effect of the falling edge, which increases with the etching depth. The spikes inside the crater seem to have little or no effect on the depth resolution.

If the etching rates for certain material compositions and plasma settings are known, it may be possible to directly correct for different etch rates in different materials according to the material concentration provided by the IBR. In-situ determination of the etching depth and thus the etching rate would be possible with the help of differential interferometric profiling [20]. These will be subject of further investigations and software improvements.

*4.2 Depth resolution*

The RSF-corrected IBR depth profiles of Al, Ga and In as well as the separate profiles of In:113 and In:115 are shown in Figure 3. As a measure for the depth resolution, we choose the approx. 2.75 nm thick indium quantum wells of the structure.

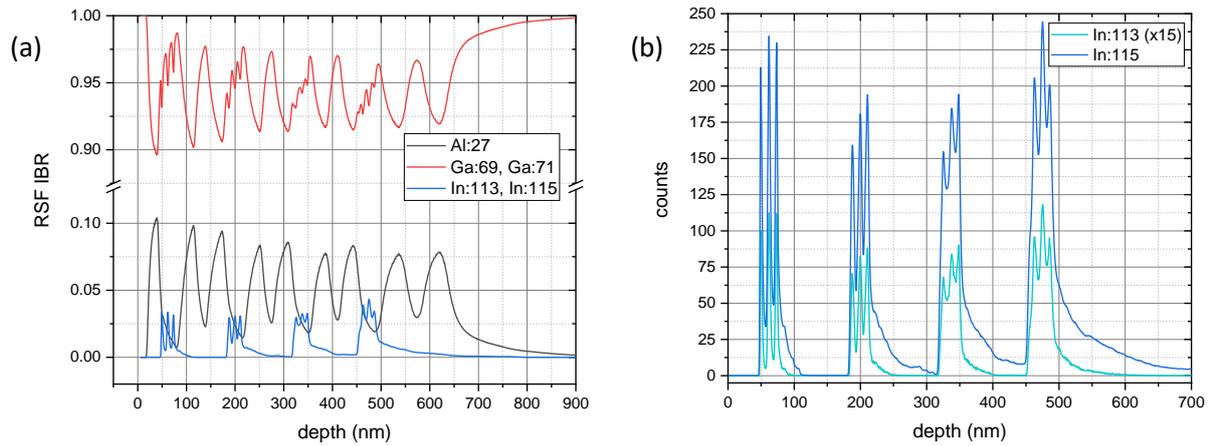

**Figure 3.** *(a) Al, Ga, In RSF-corrected IBR profile of the sample showing the 4 MQWs (55 W, 140 Pa) and (b) the corresponding PP-TOFMS signal for In:113 and In:115 ions.*

For each of the twelve QWs, both the In:113 and the In:115 isotope signal of the TOFMS (figure 3(b)) are distinctly resolved. The number of indium counts (figure 3(b)) and the IBR (figure 3(a)) increase from MQW to MQW, corresponding to the structure. A drop in the signal quality between the four MQWs can be seen in the decay of the In:115 signal after each MQW and in the level of the minima at the QBs between the individual QWs. As the etch depth of the PP-TOFMS increases, the number of In:113 and In:115 counts in the QBs increases and the decay in the signals after the MQWs is slower. This can be attributed to a deteriorating quality of the etch crater during the measurement.

*4.3 Elemental composition analysis*

*4.3.1 Determination of RSFs*

The RSFs are determined using a PP-TOFMS reference measurement (55 W, 140 Pa, figure 3) of the structure shown in figure 1.

To minimize the influence of the deteriorating etch crater, the counts for each isotope of aluminium (Al:27), gallium (Ga:69 and Ga:71) and indium (In:113 and In:115) are integrated over each MQW and each blocking layer. The integral is weighted with the width of the three QWs inside (8.25 nm for In) or the width of the blocking layers (27.5 nm for Al). The IBR is calculated out of the integral value for each MQW separately. The average In concentration in all MQWs must equal the concentrations obtained by CL/nextnano and EDX. The same holds for the Al concentration obtained by PP-TOFMS and the Al concentrations obtained by EDX and XRD. The comparison gives the RSFs for the material system and plasma settings used in this study. These are $RSF_{Ga} = 3.6$, $RSF_{Al} = 2.4$ and $RSF_{In} = 1$.

These values are used to evaluate the indium and aluminium concentrations by PP-TOFMS in further measurements.

*4.3.1 Indium concentration in InGaN-(M)QWs*

The elemental composition of the sample as a function of depth is calculated directly from the PP-TOFMS data via the RSF method. The data from three PP-TOFMS measurements is compared to STEM-EDX and CL, combined with a simulation of the optical transitions within the QWs. It is not possible to evaluate the In content with XRD measurements because the QW volume is small and the XRD data is dominated by several fringes corresponding to different periodic structures in the sample.

The indium signal quality of the MQWs decreases with increasing etching depth, as already discussed in section 4.2. Therefore, the composition of the three QWs in the top MQW is discussed individually and for the other MQWs the average compisiton of all three QWs within one MQW is analyzed.

Figure 4(a) shows that the In conctration in the single QWs is around 3.0 %–3.5 %. These values are about 3 % lower than those of the other methods, as shown in figure 4(a). If the shape of the In peaks is analyzed in more detail, only a few points of the PP-TOFMS data show a significant indium level and the slope of the signal compared to the STEM EDX data from line and overview scans is smaller. With PP-TOFMS, the spacing between the points is between 0.3 nm and 0.7 nm and the In profiles do not show a 'flat-top' shape. This could be caused by a rough bottom of the etching crater and it can be assumed that the In signal of the QW is averaged over several points. Since the mean QW thickness determined by STEM in this sample is about 2.75 nm, the In content of each QW can be integrated over several scan points and weighted with the QW thickness. The result is shown in figure 4(b). The integration leads to an In content of the QWs averaged over the three measurements of (4.8 ± 0.3) %, (6.1 ± 0.4) % and (5.5 ± 0.4) %, respectively. These values lie within the statistical error of the EDX measurement for the first two QWs and are well consistent with the average In concentration of 5.6 % obtained by CL with nextnano simulations.

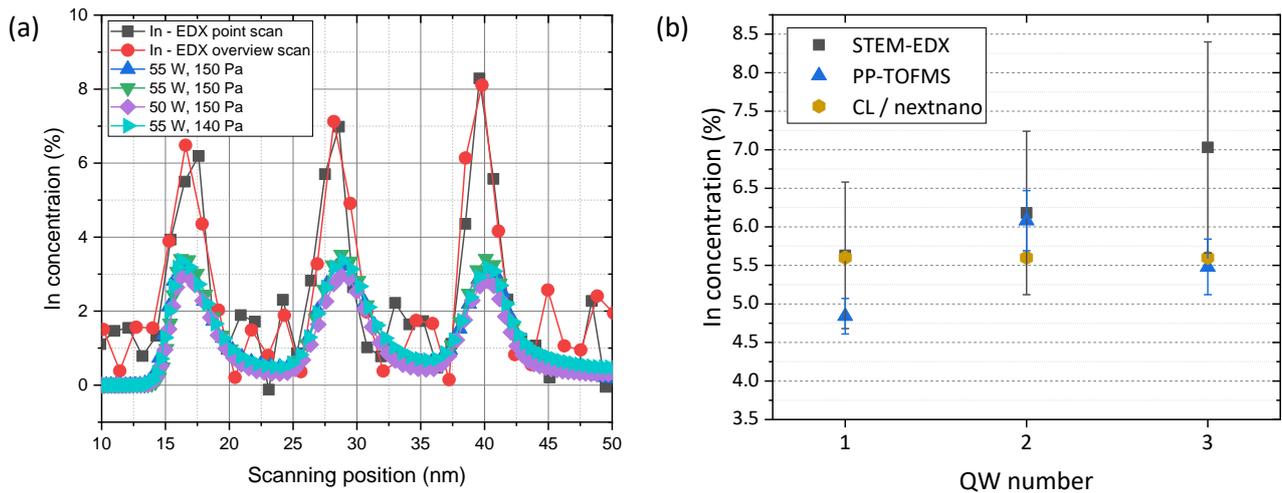

**Figure 4.** Comparison of the indium concentrations of the individual QWs for the topmost MQW obtained by different methods. (a) RSF-corrected indium concentration profiles for different PP-TOFMS measurements and STEM EDX (point- and overview-scan), (b) averaged In concentrations for the individual QWs. The PP-TOFMS data is the average of three measurements. The EDX data refer to the two highest indium contents of each QW for line scan and spot wise acquisition. The error bars represent the statistical error.

As previously mentioned, it was not possible to resolve QWs lying beyond 150 nm depth inside the crystal as precisely as the ones closer to the surface using PP-TOFMS in all measurements. Figure 3(b) shows a clear indium signal for every QW inside the MQW stack. This phenomenon might be caused by a decreasing quality of the etch crater with increasing etching time or a higher indium content in these QWs. Besides, the derivative of the signal decreases and the indium signal of the QWs no longer drops to zero in the QBs. This makes an investigation of the indium content of single QWs not feasible. However, it is possible to determine the mean indium concentration inside the MQWs by integrating the In signal over the whole MQW region and assuming a homogeneous In distribution inside the QW and a total QW thickness of 8.25 nm. The determination of the integration limits becomes more difficult with decreasing quality of the indium signal. In these cases, the derivative of the signals or the signal of the In:113 isotope (see Figure 3(b)) can be used to set the integration limits because it has a reduced tailing effect compared to the In:115. The result is shown in figure 5.

All three measurement techniques show a good correspondence between the indium concentrations obtained and differ within the range of 1 % with the exceptuin of the fourth MQW. However, the statistical error of the PP-TOFMS data is increasing slightly between MQW 1 and MQW 4, most likely caused by the decreasing quality of the etch crater. Nevertheless, individual PP-TOFMS measurements resolved all QWs very well and showed discrepancies below 1 % for the first three MQWs and around 2 % for the lowest MQW compared to EDX and CL measurements.

### *4.3.2 Aluminium concentration in AlGaN-EBLs*

The aluminium depth profile allows the determination of the aluminium concentration inside the blocking layers separating the different MQWs. Applying the same method as for indium concentration determination and assuming a thickness of 27.5 nm of each blocking layer (derived from STEM measurements), the average concentration of Al obtained by PP-TOFMS is (11.7 ± 0.5) %, whereas for STEM EDX measurements it is (11.5 ± 0.5) %. This fits well to the average Al concentration of 12 % determined from fitting an XRD ω-2θ scan around the (0002) reflex of GaN.

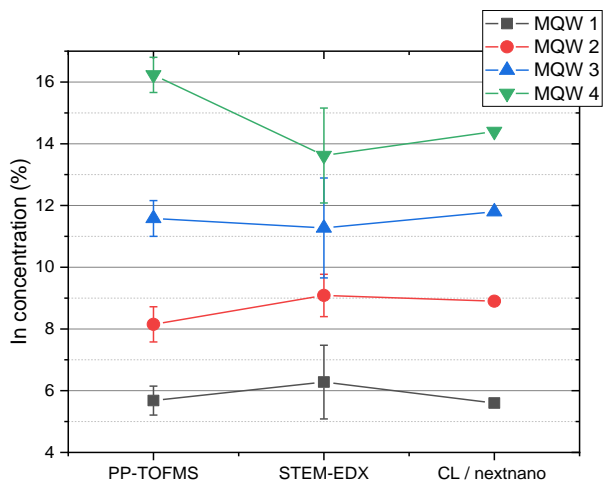

**Figure 5.** *Comparison of the indium concentrations for four MQWs with different indium content obtained by STEM-EDX, CL and PP-TOFMS. The PP-TOFMS data is the average of three measurements. The EDX data refer to the two highest indium contents of each QW for line scan and spotwise acquisition. The error bars represent the statistical error.*

4.4 XRD

The full width at half maximum (FWHM) of the (0006), (10-15) and (-1015) reflexes were determined to be 282, 171 and 313 arcsec, respectively, confirming suitable crystal quality as already observed by STEM. Fully coherent growth of the entire epilayer stack was concluded from reciprocal space maps of asymmetric (10-15) and (-1015) reflexes.

However, compositional analysis of the four different MQWs and AlGaN barriers could not be carried out reliably due to the abundance of fringes observed in e.g. a highly resolved triple-axis ω-2θ-scan of the (0002) reflex (figure 6, black curve). Different periodic structures in real space, such as the crystal lattice, but also the periodic MQW structures and the barrier spacing, give rise to interference fringes in the reciprocal space that are captured in the XRD measurement. For moderately complex samples, a simulation of the curve can be matched to the measured one by adjusting structural parameters such as alloy composition and layer thicknesses, which subsequently yields good estimates for the real values. However, to adapt a simulation to the measurement of such a complex structure as shown in figure 1, precise a-priori knowledge about the sample is necessary, which is usually not present. Thus, a conclusive analysis of the sample using XRD exclusively is not possible.

The XRD curve can be accurately simulated if, however, information on the length scales and compositions of a sample is known from either PP-TOFMS or STEM-HAADF/STEM-EDX, with only minor manual refinements to the fit required (figure 6, red curve, simulated).

**5 Conclusion**

For the Al-/Ga-/InN material system PP-TOFMS is able to resolve a few nanometer thick layers of a complex hetero-structures like MQWs down to 500 nm below the surface within a few minutes. With prior knowledge about the structure, the presented integration technique allows the determination of the relative elemental composition of indium and aluminium in III-Nitrides within about 10 rel. % of the composition obtained by EDX, CL and XRD. Exceeding an etch depth of around 400 nm, the deviation of PP-TOFMS measurements to the other measurement techniques increases, which is most likely caused by a deteriorating crater quality. Even non-ideal craters (see figure 2(a)) allow nm-thin structures like quantum wells to be resolved to depths above 500 nm (cf. figure 3) and obtain information within the aforementioned precision.

PP-TOFMS has low lateral resolution and high sample consumption compared to TEM or SIMS measurements, but the fast feedback time and low operating costs of the analysis make it an ideal tool for process development or quality control on the wafer scale. PP-TOFMS can provide complete understanding of a sample structure in question, even if standard techniques such as HRXRD fail to provide a conclusive picture on their own, or rapid feedback is needed, ruling out time-intensive STEM sample preparation and measurements.

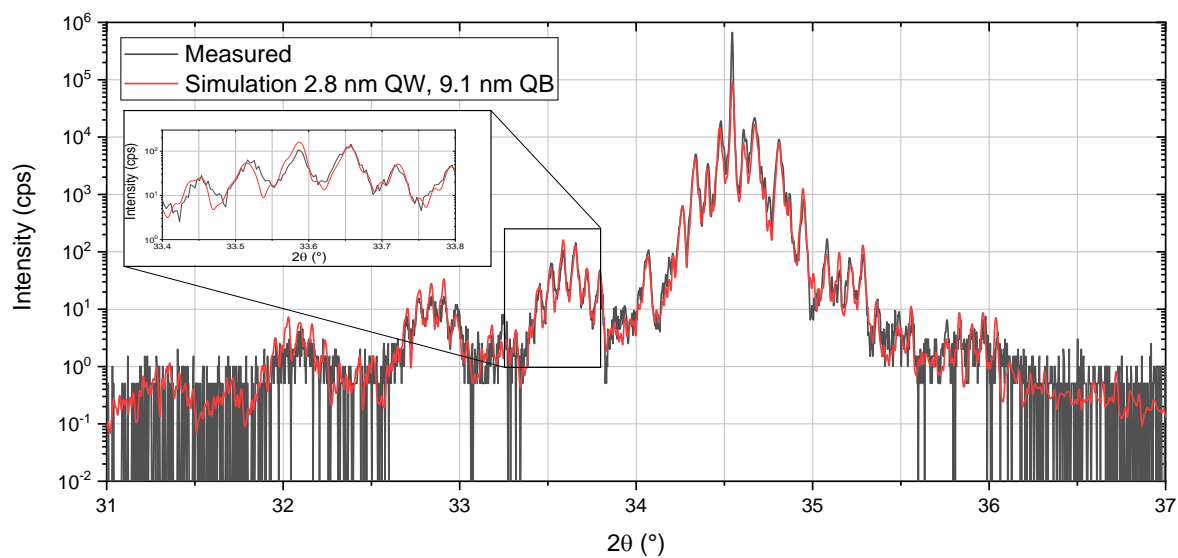

**Figure 6.** Triple-axis ω-2θ-scan around the GaN (0002) reflex (black curve) and simulated curve based on PP-TOFMS and STEM-HAAD/STEM-EDX measurements of the same structure (red curve).

The comparatively large crater and high dynamic range of PP-TOFMS might even make it applicable for the detection of background dopants or impurities in semiconductor structures, which would be of great importance for industrial applications.

**Acknowledgements:**

The authors thank Dr. Sebastien Legendre and Dr. Agnès Tempez (both Horiba Scientific, France) for valuable discussions and Aileen Michalski and Juliane Breitfelder for technical support. The work funded in part by the Lower Saxony Ministry for Science and Culture (N-MWK) and in part within the cluster of excellence 'QuantumFrontiers' funded by the German Research Society (DFG). Tobias Meyer and Michael Seibt acknowledge funding by DFG contract no. SE 560-6/1.